\documentstyle[11pt]{article}
\begin{document}
\baselineskip 5mm
\footskip 10mm \title{Indicators of Reconnection Processes and\\
Transition to Global Chaos in Nontwist Maps} \author{Susumu
Shinohara\thanks{E-mail: susumu@aizawa.phys.waseda.ac.jp} and Yoji
Aizawa\\ Department of Applied Physics, Waseda University, Tokyo
169-8555} \date{} \maketitle
\begin{abstract}
\indent Reconnection processes of twin-chains are systematically
studied in the quadratic twist map.  By using the reversibility and
symmetry of the mapping, the location of the indicator points is
theoretically determined in the phase space.  The indicator points
enable us to obtain useful information about the reconnection
processes and the transition to global chaos.  We succeed in deriving
the general conditions for the reconnection thresholds.  In addition,
a new type of reconnection process which generates shearless curves is
studied.
\end{abstract}
\section{Introduction}
In the past decades, enormous effort has been dedicated to the study
of two-dimensional area-preserving maps with the twist
condition\cite{MM}, but very few studies have been made of nontwist
maps.  Recent studies on nontwist maps have revealed that rich
properties are generated by violating the twist condition.[2-6]\\
\indent In a previous paper\cite{Susumu}, we studied the properties of
the quadratic twist map and numerically determined the critical
boundary in the two-dimensional parameter space, where the transition
to global chaos occurs.  The critical boundary has many sharp singular
structures, and their locations seem to have a one-to-one
correspondence with those of the reconnection thresholds. The
relationship between the transition to global chaos and the
reconnection processes was first pointed out by Howard and
Hohs\cite{Howard1}, but it has not yet been thoroughly investigated.\\
\indent In order to investigate the detailed structure of the critical
boundary, one needs accurate information regarding the reconnection
processes.  In this paper, we study the details of the reconnection
processes in the quadratic twist map and propose a theoretical method
to determine the reconnection thresholds.  We show that the
reversibility and symmetry of the mapping guarantee the existence of
the ``indicator points'' in the phase space.  These enable us to study
the reconnection processes systematically.  For twin-chains of period
one and period two, the reconnection thresholds have already been
determined, either exactly or approximately\cite{Howard1,Howard2}. The
method presented here reproduces results which have been previously
obtained, and it provides general conditions for the reconnection
thresholds.\\
\indent The quadratic twist map(QTM) is defined by
$$T: \left\{
\begin{array}{ll}
\theta_{n+1} = \theta_n + f_{\mu}(I_{n+1}), \\ I_{n+1} = I_{n} - K
\sin(\theta_{n}),
\end{array} \right. \eqno(1\cdot 1) $$
$$f_{\mu}(I)=2 \pi \mu-I^2, \hspace{5mm} \eqno(1\cdot2)$$
where $K$ represents the strength of the perturbation and $\mu$ the
maximum value of the twist function $f_{\mu}(I)$.  As the variable
$\theta$ is $2\pi$-periodic, $\mu$ is also periodic with period 1.
Moreover, Eq.~(1$\cdot$1) is invariant under the transformation $(K,I)
\mapsto (-K,-I)$.  Thus it is sufficient for us to consider the
parameter regions $\mu \in [-0.5,0.5)$ and $K \in [0,\infty]$.\\
\indent The mapping Eq.~(1$\cdot$1) is often called the logistic twist
map or the standard nontwist map, depending on the form of the twist
function $f_{\mu}(I)$.[2-5] In this paper, we adopt a form of
Eq.~(1$\cdot$2) for which the twist condition
$$ \frac{d f_{\mu}(I)}{d I} \ne 0 \quad \mbox{for} ~^{\forall} I
\eqno(1\cdot 3)$$
fails at $I=0$ for any value of $\mu$ in the integrable limit~($K=0$)
and for which the parameter $\mu$ itself represents the maximum value
of $f_{\mu}(I)$.\\
\indent This paper is organized as follows.  In \S 2, we derive the
location of the indicator points which play an important role in
studying the reconnection processes in the QTM and give a review of
the previous results on the transition to global chaos.  In \S 3, the
reconnection processes of even-periodic twin-chains are studied by
using the indicator points.  The reconnection thresholds are
analytically derived, and a new type of reconnection process is
analyzed.  In \S 4, we focus on the reconnection processes of
odd-periodic twin-chains, and a new numerical method to determine the
reconnection thresholds is proposed.  Section 5 contains a summary and
discussion.
\section{The indicator points of the QTM}
By reversibility, the QTM can be rewritten as $T=M_2 M_1$, where $M_1$
and $M_2$ are given by
$$ M_1: \left\{
\begin{array}{ll}
\theta^{\prime} = -\theta,\\
I^{\prime} = I - K \sin(\theta),
\end{array} \right. \eqno(2\cdot 1) $$
$$ \hspace{2cm} M_2: \left\{
\begin{array}{ll}
\theta^{\prime} = -\theta + 2\pi\mu - I^2,\\
I^{\prime} = I,
\hspace*{38mm}
\end{array} \right. \eqno(2\cdot 2) $$
and they satisfy $M_1^2=M_2^2=1$. Moreover, the mapping $T$ commutes
with the mapping $S$ (i.e., $ST=TS$) defined by
$$ S: \left\{
\begin{array}{ll}
\theta^{\prime} = \theta + \pi,\\
I^{\prime} = -I.
\end{array} \right. \eqno(2\cdot 3) $$
Previously\cite{Susumu}, we showed that when there exists only one
shearless curve in the phase space, it is an invariant set of both the
mapping $T$ and the mapping $S$.  As shown in Appendix A, the
invariant sets are given by the following sets ${\cal I}_j$~$(j=1,2)$:
$$ {\cal I}_j=\mathop{\bigcup}_{m=-\infty}^{\infty} 
\{ T^n x_j^{(m)} \}_{n=0}^{\infty}, \quad (j=1,2) 
\eqno(2\cdot 4)$$
where $x_1^{(m)}$ and $x_2^{(m)}$ are the solutions of
$$ M_1 x = R^m S x  \eqno(2\cdot 5)$$
and
$$ M_2 x = R^m S x, \eqno(2\cdot 6)$$
respectively.  Here $m$ represents an arbitrary integer, and $R$
represents the linear transformation defined by
$$ R {\theta \brack I} = {\theta - 2\pi \brack I}. \eqno(2\cdot 7) $$
The solutions of Eqs.~(2$\cdot$5) and (2$\cdot$6) are as follows:
$$ x_{1}^{(m)} = {\theta_{1}^{(m)}\brack I_{1}^{(m)}} =
{\frac{1}{2}\pi(2m-1) \brack (-1)^{m+1}~\frac{1}{2}K}, \eqno(2\cdot 8)
$$
$$ x_{2}^{(m)} = {\theta_{2}^{(m)}\brack I_{2}^{(m)}} = {\pi\mu
+\frac{1}{2}\pi(2m-1) \brack 0}. \eqno(2\cdot 9) $$
Note that $x_1^{(m)}$ depends only on $K$, while $x_2^{(m)}$ depends
only on $\mu$.  If we use the $2\pi$-periodicity of the variable
$\theta$ in the interval $(-\pi,\pi]$, it is sufficient to consider
only two terms ($m=0$ and $m=1$) in Eq.~(2$\cdot$4).  We shall call
these points, $x_j^{(m)}$ ($j=1,2$), the ``indicator points'' in the
present paper.  As will be shown in what follows, we can detect the
occurrence of the transition to global chaos and of the reconnection
of twin-chains by investigating the iterates of the indicator
points.\\
\indent When the iterates of the indicator points $x_j^{(m)}$ are
confined in a bounded region, the following two cases are possible: i)
they are confined on a shearless curve~(see Fig.~1(a)), or ii) they
are not confined on a certain KAM curve, but several robust KAM curves
surrounding them prevent global chaotic motion~(see Fig.~1(b)).  In
either case, one can say that the bounded motion of the iterates of
$x_j^{(m)}$ ensures the existence of KAM curves.  On the other hand,
the unbounded motion of the iterates of $x_j^{(m)}$ guarantees the
non-existence of KAM curves.\\
\indent As reported in Ref.~6), we have numerically determined whether
the iterates of $x_j^{(m)}$ are bounded or not for each set of $\mu$
and $K$.  The phase diagram so obtained is shown in Fig.~2, where the
iterates of $x_j^{(m)}$ are bounded in the gray region, but unbounded
in the white region.  Thus, the boundary between the gray and the
white regions displayed in Fig.~2 determines the critical boundary,
where the transition to global chaos occurs.  In numerical
calculations, we consider iterates to be bounded when the absolute
values of their $I$ components do not exceed $2$ during $10^5$ steps
of iteration~(i.e., $|I_n|<2$~for~$^{\forall}n\leq 10^5$).  These
conditions are sufficient to detect the bounded motion, because the
robust KAM curves are always localized around the indicator points
$x_j^{(m)}$, as shown in Figs.~1(a) and (b).\\
\section{Annihilation and reconnection of even-periodic twin-chains}
In this section, we consider the case where either $x_1^{(m)}$ or
$x_2^{(m)}$ is an even-periodic point with rotation number $P/Q$,
i.e.,
$$ R^{P} T^{Q} x_j^{(m)} = x_j^{(m)}, \quad (j=1,2;m\in Z)
\eqno(3\cdot 1)$$ 
where $P$ and $Q$ are relatively prime integers and let $Q$ be even.
By the symmetry of the mapping, Eq.~(3$\cdot$1) is reduced to more
simple forms~(see Appendix B):
$$ T^{Q/2} x_1^{(m)} = x_1^{(m+P)}   \eqno(3\cdot 2)$$
and
$$ T^{-Q/2} x_2^{(m)} = x_2^{(m-P)}. \eqno(3\cdot 3)$$
By solving Eqs.~(3$\cdot$2) and (3$\cdot$3), one can obtain two
independent relations between $\mu$ and $K$ parametrized by $P$ and
$Q$.  Note that these relations do not depend on $m$ at all.  The
exact iterates of $x_1^{(m)}$ and $x_2^{(m)}$ are given for $Q=2,4,6$
and $8$ in Appendix C.  The obtained $\mu$-$K$ relations are
summarized in Table I, where we put
$$ \xi(\mu,K)=1+2\cos(2\pi\mu-\frac{1}{4} K^2) \eqno(3\cdot 4)$$
and
$$ \eta(\mu,K)=1+\frac{\cos(3\pi\mu-K^2\cos^2(\pi\mu))}{\cos(\pi\mu)}. 
\eqno(3\cdot 5)$$
\indent Let us denote the $\mu$-$K$ relations derived from
Eqs.~(3$\cdot$2) and (3$\cdot$3) by $E^{(P/Q)}_1$ and $E^{(P/Q)}_2$
respectively; i.e.,
$$ E^{(P/Q)}_j=
\{(\mu,K)~|~{\hat T}_{P,Q}~x_j^{(m)}
=x_j^{(m)}\},\quad (j=1,2)\eqno(3\cdot 6)$$
where $m$ is an arbitrary integer and
$$ {\hat T}_{P,Q}=R^P T^Q. \eqno(3\cdot 7) $$
\noindent
In the $\mu$-$K$ parameter space, the two sets $E^{(P/Q)}_1$ and
$E^{(P/Q)}_2$ are described by two different curves which pass through
the trivial point $(\mu,K)=(P/Q,0)$.  These curves are shown by solid
lines in Fig.~2 for several values of $P/Q$~(those for $Q\leq 8$).
There are two types of characteristic $\mu$-$K$ curves given for each
value of $P/Q$.  We shall call these curves the right $(P/Q)$-curve or
the left $(P/Q)$-curve, depending on their relative positions,
corresponding to the right-hand side or the left-hand side in Fig.~2.
For $(P/Q)$-curves that have a crossing point, we only apply the
notion of right and left below the crossing point.  Some of the right
and the left $(P/Q)$-curves are summarized in Table II.\\
\indent Note that the $(P/Q)$-curves are quite similar to the
$(P/Q)$-bifurcation curve introduced by del-Castillo-Negrete et
al.\cite{Castillo1}, where the $(P/Q)$-bifurcation curve is defined as
the $\mu$-$K$ locus which represents the creation/annihilation points
of the twin-chains with rotation number $P/Q$.  It will be shown later
that either the right or the left $(P/Q)$-curve which we have
introduced here coincides well with the $P/Q$-bifurcation curve.\\
\indent From numerical calculations, we can find the following general
characteristics of the iterates of the indicator points $x_j^{(m)}$
and the characteristic $E^{(P/Q)}_j$ curves defined by
Eq.~(3$\cdot$6):
\begin{description}
\item[\rm (i)] For the $\mu$-$K$ parameter values given by
$E^{(P/Q)}_1$(or $E^{(P/Q)}_2$), successive iterates of $x_2^{(m)}$(or
$x_1^{(m)}$) converge to a periodic point with rotation number $P/Q$
under the mapping ${\hat T}_{P,Q}$.
\item[\rm (ii)] For the $\mu$-$K$ parameter region between the right
and the left $(P/Q)$-curves, successive iterates of both $x_1^{(m)}$
and $x_2^{(m)}$ converge to an unstable periodic point with rotation
number $P/Q$ under the mapping ${\hat T}_{P,Q}$ .
\end{description}
These properties are numerically illustrated in Fig.3 for the case of
$P/Q=-1/2$.\\
\indent It is found in many cases that twin-chains with rotation
number $P/Q$ (hereafter referred to as $(P/Q)$-twin-chains) are
created or annihilated at parameter values on the left $(P/Q)$-curve,
and that the reconnection of the $(P/Q)$-twin-chains takes place at
parameter values on the right $(P/Q)$-curve.  Thus, the left
$(P/Q)$-curve and the right $(P/Q)$-curve correspond to the
annihilation threshold and the reconnection threshold, respectively.
In Fig.~2, we can see nice agreement of the critical boundary and the
right $(P/Q)$-curves for many $P/Q$.  This is more clearly shown in
Fig.~4 for the case of $P/Q=-1/4$.  However, this correspondence does
not always hold for all $P/Q$.  For instance, in the cases of
$P/Q=-1/6$ and $P/Q=-1/8$, it is the left $(P/Q)$-curve that coincides
with the critical boundary.  This is more clearly shown in Fig.~5 for
the case of $P/Q=-1/8$.  This exceptional correspondence occurs when
the left $(P/Q)$-curve changes to correspond to the reconnection
threshold at a certain point on the curve.\\
\indent Let us illustrate this for the left $(-1/8)$-curve.  Figures
6(a)$\sim$(e) display phase space portraits for various values of $K$,
where $K$ increases along the left $(-1/8)$-curve.  At $K=0.5$,
marginally stable periodic points with rotation number $-1/8$ exist in
the phase space, as shown in Fig.6(a).  At $K=K_1\simeq 0.519$, the
periodic point becomes unstable and gives birth to two elliptic
points, as shown in Fig.~6(b).  We refer to the chain of these
periodic points as the primary chain.  At $K=K_2\simeq 0.5492$,
another pair of periodic points with rotation number $-1/8$ is created
on both sides of the primary chain via a saddle-node(SN) bifurcation.
Figure 6(c) exhibits the phase space portrait at $K=0.552$, where two
chains of SN pairs appear, in addition to the primary chain mentioned
above.  We refer to these chains of SN pairs as the secondary chains.
Moreover, at $K=K_3\simeq 0.561$, the two separatrices of the primary
chain and those of the secondary chains merge, as shown in Fig.~6(d).
As a result of the reconnection process induced by the merging of the
primary and secondary chains, vortex pairs and two Poincar\'e-Birkhoff
chains come to appear, as shown in Fig.~6(e).\\
\indent To summarize, the left $(-1/8)$-curve corresponds to the
annihilation threshold of the primary chain for $K<K_1$.  While, for
$K>K_3$, it corresponds to the reconnection threshold of the vortex
pairs that are formed via the reconnection process of the primary and
secondary chains.\\
\section{Reconnection of odd-periodic twin-chains}
\indent In this section we characterize the reconnection of
odd-periodic twin-chains by using the indicator points.  At a
reconnection threshold of the $(P/Q)$-twin-chains, successive iterates
of both $x_1^{(m+P)}$ and $T^{\frac{Q-1}{2}} x_2^{(m)}$ approach the
same hyperbolic periodic point of the reconnecting twin-chains under
the mapping $\hat{T}_{P,Q}$.  Figure 7(a) corresponds to the case of
the $(-1/3)$-twin-chains at the reconnection threshold.  Figure 7(b)
is the magnification of the phase space near the lower hyperbolic
periodic point of the $(-1/3)$-twin-chains, where ${\hat
T}_{-1,3}^n~x_1^{(0)}$ and ${\hat T}_{-1,3}^n~T^{1} x_2^{(1)}$
approach the same hyperbolic periodic point as $n$ increases.  We have
numerically confirmed this property for all odd-periodic twin-chains
with period $Q \leq 9$.  This property seems to hold even at a large
$K$-value, as shown in Fig.~8, where we can still observe monotonic
convergence in spite of strong chaos around a hyperbolic periodic
point.\\
\indent Consider the quantity $\delta X_n$ given by
$$ \delta X_n = ||~\hat{T}_{P,Q}^n~x_1^{(m+P)}-
\hat{T}_{P,Q}^n~T^{\frac{Q-1}{2}}x_2^{(m)} ||, \eqno(4\cdot 1)$$
where $m$ is an arbitrary integer and $||\cdot||$ represents the
Euclidian norm.  At the reconnection threshold, $\delta X_n$ is
expected to decrease exponentially as
$$ \delta X_n \propto \lambda^n, \eqno(4\cdot 2)$$
where $\lambda$ represents the smaller~(stable) eigenvalue of the
tangent map of the mapping $T^Q$ evaluated at the hyperbolic periodic
point of the $(P/Q)$-twin-chains.  Numerical evidence for the
exponential scaling regime of Eq.~(4$\cdot$2) at the reconnection
threshold of the $(-1/3)$-twin-chains is shown in Fig.~9 for various
$K$-values.  There appear large deviations from the scaling regime
when the number of iterations $n$ becomes extremely large, because the
monotonic convergence of $\delta X_n$ is violated by heteroclinic
chaos near the hyperbolic periodic point.  However, as shown in Table
III, the decay rates in the scaling regime are in good agreement with
the theoretically estimated values of $\ln(\lambda)$.\\
\indent On the basis of the remarkable convergent property mentioned
above, the determination of the reconnection threshold is reduced to
the problem of finding the parameter values where $\delta X_N\leq
\epsilon$ holds for a given small value $\epsilon$ with a certain
large $N$.  The merit of this computational criterion is that it does
not require any additional knowledge about the detailed structure of
twin-chains, such as the location of the periodic point and its
stability.  Applying this criterion, we numerically determined the
reconnection thresholds for odd-periodic twin-chains with the period
$Q \leq 9$, which are indicated by solid lines for $Q$=odd in Fig.~2.
The results based on the above technique seem to reproduce the
reconnection thresholds.  This is illustrated more clearly in Fig.~10
for the case of the $(-1/3)$-twin-chains.
\section{Summary and discussion}
We have studied in detail the reconnection processes of twin-chains in
the QTM by investigating the iterates of the indicator points
$x_j^{(m)}$ which belong to the invariant sets of both the mapping $T$
and the mapping $S$.  For the case of even-periodic twin-chains, we
introduced $(P/Q)$-curves in the parameter space, on which either the
annihilation or the reconnection of the $(P/Q)$-twin-chains occurs.
We succeeded in deriving the $(P/Q)$-curves analytically for $Q \leq
8$.  One can also derive $(P/Q)$-curves with larger value of $Q$ by
solving Eqs.~(3$\cdot$2) and (3.3).  On the other hand, for the case
of odd-periodic twin-chains, we propose a numerical method to
determine the reconnection threshold.  The method is based on the
observation that the successive iterates of the indicator points
converge to the same hyperbolic periodic point at the reconnection
threshold.  Numerical results clearly show that the critical boundary
is described well by the reconnection thresholds of twin-chains.\\
\indent Finally, we briefly discuss the reconnection process between
the primary and the secondary twin-chains which was found for the
$(-1/8)$-twin-chains.  The variety of reconnecting twin-chains has
been studied in cubic and quartic twist maps\cite{Howard2}.  For
example, cubic twist maps exhibit the reconnection of three island
chains with the same rotation number.  This is a natural consequence
of the fact that the cubic twist function generically has two extrema
and that each extremum generates an inherent reconnection process.
However, it is surprising that multiple reconnection processes occur
simultaneously even in the QTM whose twist function has only one
extremum.  The occurrence of the secondary twin-chains inevitably
ensures the existence of multiple shearless curves in the phase space,
neither of which passes through the indicator points $x_j^{(m)}$(see
Figs.~6(c) and (e)).  Our result suggests that shearless curves can be
created by nonintegrable nonlinear perturbations.  This problem will
be discussed elsewhere.
\newpage
\appendix
\section*{Appendix A\quad The Invariant Sets ${\cal I}_{j}$~$(j=1,2)$}
It is trivial to show that ${\cal I}_j~(j=1,2)$ given by
Eq.~(2$\cdot$4) is an invariant set of $T$.  Here we show that ${\cal
I}_j$ is invariant under the transformation $S$, i.e, $S {\cal
I}_j={\cal I}_j ~(j=1,2)$.  In order to do this, we show that $ST^n
x_j^{(m)}$ is also an element of ${\cal I}_j$ as follows:
$$ ST^n x_1^{(m)}=T^n S x_1^{(m)}=T^n R^{-m}M_1 x_1^{(m)} =T^n M_1
x_1^{(3m)}$$
$$=T^{n-1}M_2 x_1^{(3m)}=T^nx_1^{(1-3m)},\hspace*{7mm}
\eqno(\mbox{A}\cdot 1)$$
$$ ST^n x_2^{(m)}=T^n S x_2^{(m)}=T^n R^{-m}M_2 x_2^{(m)} =T^n M_2
x_2^{(3m)}$$
$$=T^{n+1}M_1 x_2^{(3m)}=T^nx_2^{(1-3m)},\hspace*{7mm}
\eqno(\mbox{A}\cdot 2)$$
where we have used the relation $ST=TS$ in the first equalities, and
Equations (2$\cdot$5) and (2$\cdot$6) in the second equalities.
Moreover, we use the following relations in the last equalities:
$$ M_2 x_1^{(m)}=T x_1^{(1-m)}, \eqno(\mbox{A}\cdot 3)$$
$$ M_1 x_2^{(m)}=T^{-1} x_2^{(1-m)}. \eqno(\mbox{A}\cdot 4)$$
\section*{Appendix B\quad Derivation of Eqs.(3$\cdot$2) and (3$\cdot$3)}
Assume that $x_j^{(m)}~(j=1,2;m \in Z)$ given by Eqs.~(2$\cdot$8) and
(2$\cdot$9) are even-periodic points with rotation number $P/Q$; i.e.,
$$ R^P T^Q x_j^{(m)}=x_j^{(m)}.~(j=1,2) \eqno(\mbox{B}\cdot 1)$$
The property of the involutions $M_1$ and $M_2$ implies
$$ M_1 T^{Q/2} = T^{-Q/2} M_1, \quad M_2 T^{-Q/2} = T^{Q/2} M_2.
\eqno(\mbox{B}\cdot 2)$$
Therefore we have
$$ M_1 T^{Q/2} x_1^{(m)} = T^{-Q/2} M_1 x_1^{(m)} = T^{-Q/2} R^{m} S
x_1^{(m)} = R^{m} S T^{-Q/2} x_1^{(m)}\eqno(\mbox{B}\cdot 3)$$ and
$$ M_2 T^{-Q/2} x_2^{(m)} = T^{Q/2} M_2 x_2^{(m)} = T^{Q/2} R^{m} S
x_2^{(m)} = R^{m} S T^{Q/2} x_2^{(m)}, \eqno(\mbox{B}\cdot 4)$$
where we have used Eqs.~(2$\cdot$5) and (2$\cdot$6) in the second
equalities and the relation $ST=TS$ in the third equalities.\\
\indent On the other hand, Eq.~(B$\cdot$1) implies
$$ T^{-Q/2}x_j^{(m)}= R^{P} T^{Q/2}x_j^{(m)},
\quad(j=1,2)\eqno(\mbox{B}\cdot 5)$$
$$ T^{Q/2}x_j^{(m)}= R^{-P}
T^{-Q/2}x_j^{(m)}. \quad(j=1,2)\eqno(\mbox{B}\cdot 6)$$
Substituting (B$\cdot$5) and (B$\cdot$6) into (B$\cdot$3) and
(B$\cdot$4), respectively, we have
$$ M_1 T^{Q/2}x_1^{(m)}= R^{m+P} S T^{Q/2}x_1^{(m)}
\eqno(\mbox{B}\cdot 7)$$ and
$$ M_2 T^{-Q/2}x_2^{(m)}= R^{m-P} S T^{-Q/2}
x_2^{(m)}. \eqno(\mbox{B}\cdot 8)$$
Since these equations are equivalent to Eqs.~(2$\cdot$5) and
(2$\cdot$6), the solutions are given by
$$ T^{Q/2} x_1^{(m)} = x_1^{(m+P)}, \eqno(\mbox{B}\cdot 9)$$
$$ T^{-Q/2} x_2^{(m)} = x_2^{(m-P)}, \eqno(\mbox{B}\cdot 10)$$
respectively.
\section*{Appendix C\quad Exact Iterations of $x_1^{(m)}$ and $x_2^{(m)}$}
The exact expressions for $T^n x_1^{(m)}$~$(n=0,\pm1,\pm2)$ are given
by
\vspace*{-5mm}
\begin{flushleft}
\begin{minipage}[t]{11.5cm}
\begin{eqnarray*}
T^{-2} x_1^{(m)} &=& {
\frac{\pi}{2}(2m-1)-4\pi\mu+\frac{1}{4}K^2(1+\xi(\mu,K)^2) \brack
(-1)^{m+1}\frac{1}{2}
K\{\xi(\mu,K)+2\cos(4\pi\mu-\frac{1}{4}K^2(1+\xi(\mu,K)^2))\} },\\
T^{-1} x_1^{(m)} &=& { \frac{\pi}{2}(2m-1)-2\pi\mu+\frac{1}{4} K^2
\brack (-1)^{m+1}\frac{1}{2}K \xi(\mu,K) },\\
x_1^{(m)} &=& { \frac{\pi}{2}(2m-1) \brack (-1)^{m+1}\frac{1}{2}K },\\
T x_1^{(m)} &=& { \frac{\pi}{2}(2m-1)+2\pi\mu-\frac{1}{4}K^2 \brack
(-1)^m\frac{1}{2}K },\\
T^2 x_1^{(m)} &=& {
\frac{\pi}{2}(2m-1)+4\pi\mu-\frac{1}{4}K^2(1+\xi(\mu,K)^2) \brack
(-1)^m \frac{1}{2}K \xi(\mu,K) },
\end{eqnarray*}
\end{minipage}
\begin{minipage}[t]{3cm}
\begin{flushright}
\vskip 6mm $(\mbox{C}\cdot 1)$ \vskip 7mm $(\mbox{C}\cdot 2)$ \vskip
7mm $(\mbox{C}\cdot 3)$ \vskip 7mm $(\mbox{C}\cdot 4)$ \vskip 7mm
$(\mbox{C}\cdot 5)$
\end{flushright}
\end{minipage}
\end{flushleft}
where $\xi(\mu,K)$ is given by Eq.~(3$\cdot$4).\\ 
\indent While the exact expressions for $T^n
x_2^{(m)}$~$(n=0,\pm1,\pm2)$ are given by
\vspace*{-5mm}
\begin{flushleft}
\begin{minipage}[t]{11.5cm}
\begin{eqnarray*}
T^{-2} x_2^{(m)} &=& { \frac{\pi}{2}(2m-1)-3\pi\mu+K^2 \cos^2(\pi\mu)
\brack (-1)^{m+1} K \cos(\pi\mu) \eta(\mu,K) },\\
T^{-1} x_2^{(m)} &=& { \frac{\pi}{2}(2m-1)-\pi\mu \brack (-1)^{m+1} K
\cos(\pi\mu) },\\
x_2^{(m)} &=& { \frac{\pi}{2}(2m-1)+\pi\mu \brack 0 },\\
T x_2^{(m)} &=& { \frac{\pi}{2}(2m-1)+3\pi\mu-K^2\cos^2(\pi\mu) \brack
(-1)^m K \cos(\pi\mu) },\\
T^2 x_2^{(m)} &=& { \frac{\pi}{2}(2m-1)+5\pi\mu-K^2\cos^2(\pi\mu)
(1+\eta(\mu,K)^2) \brack (-1)^m K \cos(\pi\mu) \eta(\mu,K) },
\end{eqnarray*}
\end{minipage}
\begin{minipage}[t]{3cm}
\begin{flushright}
\vskip 6mm $(\mbox{C}\cdot 6)$ \vskip 7mm $(\mbox{C}\cdot 7)$ \vskip
7mm $(\mbox{C}\cdot 8)$ \vskip 7mm $(\mbox{C}\cdot 9)$ \vskip 7mm
$(\mbox{C}\cdot 10)$
\end{flushright}
\end{minipage}
\end{flushleft}
where $\eta(\mu,K)$ is given by Eq.(3$\cdot$5).
\newpage
\section*{Tables}
\vskip 15mm
\begin{center}
Table I.\quad The $\mu$-$K$ relations derived from Eqs.~(3$\cdot$2)
and (3$\cdot$3).
\vskip 4mm
\begin{tabular}{c|c|c|c}
\hline \hline \multicolumn{1}{c|}{$Q$}& \multicolumn{1}{c|}{$P$} &
\multicolumn{1}{l|}{The $\mu$-$K$ relation obtained from} &
\multicolumn{1}{l}{The $\mu$-$K$ relation obtained from}\\ & &
\multicolumn{1}{r|}{$ R^P T^{Q} x_1^{(m)} = x_1^{(m)}$} &
\multicolumn{1}{r}{$ R^P T^{Q} x_2^{(m)} = x_2^{(m)}$}\\ \hline $2$ &
$-1$ & \multicolumn{1}{l|}{$ 8\pi(\mu-P/Q)=K^2 $} &
\multicolumn{1}{l}{$ \mu=-1/2$} \\ $4$ & $\pm 1$ &
\multicolumn{1}{l|}{$ 8\pi(\mu-P/Q)=K^2 $} & \multicolumn{1}{l}{$
4\pi(\mu-P/Q)=K^2 \cos^2(\pi\mu)$}\\ $6$ & $\pm 1$ &
\multicolumn{1}{l|}{$ 24\pi(\mu-P/Q)=K^2(2+\xi(\mu,K)^2) $} &
\multicolumn{1}{l}{$ 3\pi(\mu-P/Q)=K^2 \cos^2(\pi\mu)$} \\ $8$ & $\pm
3,\pm 1$ & \multicolumn{1}{l|}{$ 16\pi(\mu-P/Q)=K^2(1+\xi(\mu,K)^2) $}
& \multicolumn{1}{l} {$
8\pi(\mu-P/Q)=K^2\cos^2(\pi\mu)(2+\eta(\mu,K)^2)$} \\ \hline
\end{tabular}
\end{center}
\vskip 15mm
\begin{center}
Table II.\quad The relative location of the two curves\\ corresponding
to $E_1^{(P/Q)}$ and $E_2^{(P/Q)}$.
\vskip 4mm
\begin{tabular}{c|c|c|c}
\hline \hline $Q$ & $P$ & left & right\\ \hline $2$ & $-1$ & $E_2$ &
$E_1$\\ \hline $4$ & $-1$ & $E_1$ & $E_2$\\ & $+1$ & $E_2$ & $E_1$\\
\hline $6$ & $-1$ & $E_2$ & $E_1$\\ & $+1$ & $E_2$ & $E_1$\\ \hline
$8$ & $-3$ & $E_1$ & $E_2$\\ & $-1$ & $E_1$ & $E_2$\\ & $+1$ & $E_2$ &
$E_1$\\ & $+3$ & $E_2$ & $E_1$\\ \hline
\end{tabular}
\end{center}
\vskip 15mm
\begin{center}
Table III.\quad Slopes of the scaling regime in Fig.9\\ and
theoretically estimated values of $\ln(\lambda)$.
\vskip 4mm
\begin{tabular}{ccc}
\hline \hline $K$ & slope & $\ln(\lambda)$ \\ \hline $0.40$ & $-0.140$
& $-0.142$ \\ $0.60$ & $-0.320$ & $-0.325$ \\ $0.85$ & $-0.640$ &
$-0.668$ \\ \hline
\end{tabular}
\end{center}
\newpage
\section*{Figure captions}
\begin{description}
\item[Fig.~1.] Two types of bounded motions.  (a) $(\mu,K)$ =
$(-0.366328694,0.80)$, (b) $(\mu,K)$ = $(0.089259,0.28)$.  In both
figures, 50 successive iterates of $x_1^{(1)}$ and $x_2^{(0)}$ under
the mapping $T$ are plotted by solid circles~($\bullet$) and
crosses~($\times$), respectively.
\item[Fig.~2.] Phase diagram of the quadratic twist map.  The iterates
of the indicator points $x_j^{(m)}$ are bounded in the gray region but
unbounded in the white region.  The boundary between the gray and the
white regions corresponds to the critical boundary, where the
transition to global chaos occurs.  The solid lines represent the
$(P/Q)$-curves for even-periodic twin-chains with period $Q \leq 8$
and the reconnection thresholds for odd-periodic twin-chains with
period $Q \leq 9$.
\item[Fig.~3.] Phase space portraits and successive iterates of
$x_1^{(1)}$ and $x_2^{(1)}$ under the mapping ${\hat
T}_{-1,2}=R^{-1}T^2$, where solid circles~($\bullet$) represent the
iterates of $x_1^{(1)}$, and crosses~($\times$) represent those of
$x_2^{(1)}$.  (a) At $(\mu,K)$ = $(-0.5,0.5)$ $\in$ $E_2^{(P/Q)}$,
$x_2^{(1)}$ is a marginally stable periodic point, and successive
iterates of $x_1^{(1)}$ converge to $x_2^{(1)}$ under the mapping
${\hat T}_{-1,2}$.  (b) At $(\mu,K)$ = $(-0.490052816,0.5)$ $\in$
$E_1^{(P/Q)}$, $x_1^{(1)}$ is an unstable periodic point, and
successive iterates of $x_2^{(1)}$ converge to $x_1^{(1)}$ under the
mapping ${\hat T}_{-1,2}$.  (c) At $(\mu,K)=(-0.495,0.5)$, successive
iterates of both $x_1^{(1)}$ and $x_2^{(1)}$ converge to an unstable
periodic point under the mapping ${\hat T}_{-1,2}$.
\item[Fig.~4.] Magnification of the phase diagram and the right and
left $(-1/4)$-curves.
\item[Fig.~5.] Magnification of the phase diagram and the right and
left $(-1/8)$-curves.
\item[Fig.~6.] The reconnection process of the primary and secondary
chains with rotation number $P/Q$=$-1/8$. The values of $\mu$ and $K$
increase along the left $(-1/8)$-curve as follows: (a) $(\mu,K)$ =
$(-0.084975,0.50)$, (b) $(\mu,K)$ = $(-0.079063,0.53)$, (c) $(\mu,K)$
= $(-0.074355,0.552)$, (d) $(\mu,K)$ = $(-0.072334,0.561)$, (e)
$(\mu,K)$ = $(-0.070258,0.57)$.
\item[Fig.~7.] (a) Phase space portraits of the $(-1/3)$-twin-chains
at a reconnection threshold $(\mu,K)$ = $(-0.32681945,0.4)$. (b)
Magnification of the box in Fig.~7(a). 90 successive iterates of
$x_1^{(0)}$ and $T~x_2^{(1)}$ under the mapping ${\hat T}_{-1,3}$ are
also shown as solid circles~($\bullet$) and crosses~($\times$),
respectively.
\item[Fig.~8.] (a) Phase space portraits of the $(-1/3)$-twin-chains
at a reconnection threshold $(\mu,K)$ = $(-0.30159859,0.85)$. (b)
Magnification of the box in Fig.~8(a). 12 successive iterates of
$x_1^{(0)}$ and $T~x_2^{(1)}$ under the mapping ${\hat T}_{-1,3}$ are
also shown as solid circles~($\bullet$) and crosses~($\times$),
respectively.
\item[Fig.~9.] Numerical evidence for the exponential decrease of $\delta
X_n$ at a reconnection threshold for $K=0.4$, $K=0.6$ and $K=0.85$.
\item[Fig.~10.] Magnification of the phase diagram and the reconnection
threshold for the $(-1/3)$-twin-chains.
\end{description}

\begin{thebibliography}{99}
\bibitem{MM}{\it Hamiltonian Dynamical Systems}, Ed.~R.~S.~MacKay and
J.~D.~Meiss~(Adam Hilger, London, 1987).
\bibitem{Howard1}J.~E.~Howard and S.~M.~Hohs, Phys.~Rev. {\bf A29}
(1984), 418.
\bibitem{Howard2}J.~E.~Howard and J.~Humpherys, Physica {\bf D80}
(1995), 256.
\bibitem{Castillo1}D.~del-Castillo-Negrete, J.~M.~Greene and
P.~J.~Morrison, Physica {\bf D91} (1996), 1.
\bibitem{Castillo2}D.~del-Castillo-Negrete, J.~M.~Greene and
P.~J.~Morrison, Physica {\bf D100} (1997), 311.
\bibitem{Susumu}S.~Shinohara and Y.~Aizawa, Prog.~Theor.~Phys. {\bf
97} (1997), 379.
\end{thebibliography}
\end{document}